\documentclass[%
 reprint,
amsmath,
amssymb,
aps,
pra,
]{revtex4-1}
\usepackage{graphicx}
\usepackage{dcolumn}
\usepackage{bm}
\usepackage{subfigure}

\begin{document}

\preprint{APS/123-QED}

\title{Continuous-variable controlled-$Z$ gate using an atomic ensemble}

\author{Ming-Feng Wang}
\author{Nian-Quan Jiang}
\author{Qing-Li Jin}
\author{Yi-Zhuang Zheng}
\email{yzzheng@wzu.edu.cn}
\affiliation{School of Physics $\&$ Electronics Information Engineering, Wenzhou University, Wenzhou 325027, People's Republic of China}%
\date{\today}

\begin{abstract}
The continuous-variable controlled-$Z$ gate is a canonical two-mode gate for universal continuous-variable quantum computation. It is considered as one of the most fundamental continuous-variable quantum gates. Here we present a scheme for realizing continuous-variable controlled-$Z$ gate between two optical beams using an atomic ensemble. The gate is performed by simply sending the two beams propagating in two orthogonal directions twice through a spin-squeezed atomic medium. Its fidelity can run up to one if the input atomic state is infinitely squeezed. Considering the noise effects due to atomic decoherence
and light losses, we show that the observed fidelities of the scheme are still quite high within presently available techniques.\\
\begin{description}
\item[PACS numbers]
03.67.Lx, 03.67.Bg, 42.50.Dv
\end{description}
\end{abstract}

\maketitle


\section{introduction}
Quantum computation (QC) is one of the most fascinating and fruitful area of research. It strives to utilize the principles of
quantum mechanics to improve the efficiency of computation. In recent years, with the discoveries of many quantum algorithms \cite{Ref1,PhysRevLett.79.325}, QC has been shown to work faster than any known classic computation. Though originally based upon discrete variables (DV), QC over continuous variables (CV) has also been addressed by Lloyd and Braunstein \cite{PhysRevLett.82.1784}.
In analogy to DV QC based on gate operations, universal CV QC can be carried out by executing a finite set of CV quantum logic gates, including displacement gate, shearing gate, cubic phase gate, and controlled-$Z$ ($C_Z$) gate \cite{PhysRevA.79.062318}.

 CV $C_Z$ gate [also called as quantum nondemolition (QND) gate] is a canonical two-mode gate, which is a CV analog of a two-qubit CPHASE gate. Like the DV CPHASE gate, CV $C_Z$ gate is considered as one of the most fundamental CV
quantum gates.
Nowadays, much effort has been devoted
to the realization of such gate in many physical systems, especially in optical systems \cite{NATRUE10,PhysRevLett.57.2473,OC1,PhysRevA.42.2995,PhysRevLett.66.1418,PhysRevLett.62.28,PhysRevA.71.055801,PhysRevA.71.042308,PhysRevLett.101.250501}. However, by now optical CV CZ gate is still experimentally challenging.
Initial approaches to realize optical QND interactions are based on nonlinear optical systems \cite{PhysRevLett.57.2473,OC1,PhysRevA.42.2995,PhysRevLett.66.1418,PhysRevLett.62.28}.
$C_Z$ gate is performed by simply sending the beams through an optical crystal or an optical fiber.
These approaches, however, are hampered by the weak nonlinearities in the nonlinear media. Although the nonlinearities can be enhanced by embedding the crystal inside a cavity or using a long fiber, such enhancement techniques, on the other hand, make it difficult to inject the light states into the system and cause large loss rates. Recently, an alternative approach was proposed to circumvent cumbersome nonlinear interactions by using only linear optical elements and off-line squeezed light beams \cite{PhysRevA.71.042308}. This approach, however, requires efficient homodyne detection techniques and accurate feed forward control, which  poses a challenge to the experimental realization \cite{PhysRevLett.101.250501}.
In recent years, with the emergence of one-way QC \cite{PhysRevLett.97.110501}, the interest in theoretical and experimental realization of optical $C_Z$ gate is further fuelled.

 One-way QC is a new form of QC which eliminates unitary evolution and relies solely on adaptive measurements on a suitably prepared multiparty entangled state. This model is quite attractive because local projective measurements are often easier to implement than unitary evolution. Most of the challenging work of QC are then converted into the problem of creating the  multiparty entangled state---the so-called \emph{cluster state}.
Cluster state in the CV regime is a multimode squeezed Gaussian state and has been proved universal resources for CV one-way QC. To date, several methods have been proposed to construct such state. One of them is called the \emph{canonical method} \cite{PhysRevA.73.032318}, which involves offline single-mode squeezers and $C_Z$ gates. Although this method is conceptually simple, it is not very practical because of the experimental challenges associated with $C_Z$ gates. Shortly after this method, another method, \emph{linear-optics method} \cite{PhysRevA.76.032321}, has also been proposed to eliminate the need for $C_Z$ gates. Any desired CV cluster state can be created by using only offline-squeezing and linear optics. This method, however, suffers from drawback that effects its scalability. Recently, Menicucci \emph{et al.} proposed the \emph{single-QND-gate method} \cite{PhysRevLett.104.250503} which reintroduces the $C_Z$ gates. Although this method has many distinct advantages, i.e. it saves the resources needed greatly and eliminates the need for long-time coherence of a large cluster state, still the inefficiency of the experimental realization of $C_Z$ gates casts it into the shade. Apparently, the only way to repolish the $C_Z$-based methods (and thus the CV one-way QC) is to devise many new schemes which enable us to implement the $C_Z$ gate more efficiently.

In this paper, we propose a simple and practical scheme to realize optical $C_Z$ gate using a free-space macroscopic atomic ensemble. We show that $C_Z$ gate between two optical beams can be performed by simply sending them perpendicularly to each other twice through a spin-squeezed atomic ensemble, see Fig. 1(a). The fidelities obtained in the scheme depend solely on the degree of the atomic squeezing. The more the amount of squeezing input, the higher the fidelity obtain.
Near-unity fidelity can be achieved under the condition that the atomic state is infinitely squeezed. Unlike previous measurement-based schemes \cite{PhysRevLett.101.250501,PhysRevA.71.042308,*PhysRevA.80.050303}, our scheme requires neither homodyne detection nor feedforward control during the gate operation, which greatly simplifies its experimental implementation. Within the presently experimentally available parameters, we find that the observed fidelities are quite high even with room-temperature atomic vapors.

 \makeatletter
    \newcommand{\rmnum}[1]{\romannumeral #1}
    \newcommand{\Rmnum}[1]{\expandafter\@slowromancap\romannumeral #1@}
    \makeatother
The remainder of this paper is organized as follows. In sec. \Rmnum{2}
  we first review some basic theories, and then give details of $C_Z$ operation based on an atomic ensemble.
  In sec. \Rmnum{3} we will consider the noise effects. After that, the experimental feasibility of the scheme is also discussed.
 Finally, sec. \Rmnum{4} contains brief conclusions.
 \begin{figure}[tp]
\centering
\includegraphics[scale=0.8]{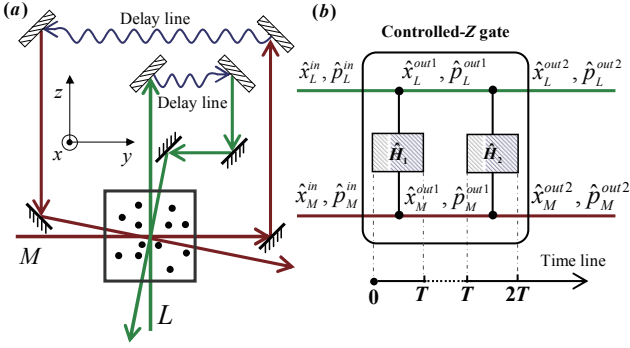}
\caption{(Color online) (a) Scheme setup for realization of optical $C_Z$ gate in an atomic ensemble. Two light beams $L$ and $M$ simultaneously enter a spin-squeezed atomic ensembles. The outgoing beams then pass through a delay line. After the whole of the beams run through the atomic sample, they enter the ensemble again. When the beams exit, a $C_Z$ gate between $L$ and $M$ is performed. (b) Schematic chart of the $C_Z$ gate. The two interactions do not overlap in the time line. After them, the gate is performed.}
\end{figure}
\section{BASIC THEORY AND CONTROLLED-Z Gate}
\subsection{Basic theory}

For a two-mode two-party system described by the quadrature operators $\hat x_{1,2}$ and $\hat p_{1,2}$ satisfying the commutation relation $[\hat x_j,\hat p_k]=i\delta_{jk}$, the QND-gate coupling Hamiltonian can be written as: $\hat H_{QND}=\hbar \chi
\hat p_1\hat p_2$, where $\chi$ is the coupling coefficient. In the Heisenberg picture, one may deduce the ideal QND input-output relations for both position and momentum operators
\begin{eqnarray}
  \hat x_1^{out} = \hat x_1^{in} + G\hat p_2^{in},~~~\hat p_1^{out} = \hat p_1^{in}, \hfill \nonumber\\
  \hat x_2^{out} = \hat x_2^{in} + G\hat p_1^{in},~~~\hat p_2^{out} = \hat p_2^{in}, \label{equ:0}
\end{eqnarray}
where $G=\chi t$ is the gain of the interaction, and $t$ represents the interaction time. For nonzero $G$, these equations imply that the two sub-systems become Gaussian entangled \cite{RevModPhys.77.513}. Such entanglement has been widely used in quantum information processing \cite{PhysRevLett.85.5639,PhysRevLett.85.5643,PhysRevLett.91.060401}. Specifically, if we put the gain of the interaction $G=1$, then we obtain the CV $C_Z$ gate as desired.

To realize optical $C_Z$ gate in an atomic ensemble, let us first investigate the interaction between light and atoms.
Consider a cell filled with a large number of atoms interacting with a light pulse traveling along the $z$ direction. The atoms in the cell are initially prepared in a coherent spin state, i.e. a fully polarized state along the $x$ axis. As a result, we may treat the $x$ component of the collective spin as a $c$ number, that is $\hat J_x$ by $J_x$. In this case, we can map the transverse spin components into dimensionless canonical variables $(\hat x_a,\hat p_a)=(\hat J_y,\hat J_z)/\sqrt{J_x}$ obeying $[\hat x_a,\hat p_a]=i$. The light pulse interacting with atoms is also linearly polarized along the $x$ axis. Similarly, we may define the optical canonical operators as $(\hat x_{ph}(t),\hat p_{ph}(t))=(\hat S_y(t),\hat S_z(t))/\sqrt{S_x}$, which satisfy the commutation relation $[\hat x_{ph}(t),\hat p_{ph}(t)]=i\delta(t-t')$ and have the dimensions of $1/\sqrt{time}$, where $\hat S_i$ (with $i\in\{x,y,z\}$) denotes the time dependent Stokes vector component. Under the condition that the frequency of the beam was tuned far off resonance with atomic transition \cite{EPL1}, the interaction of light with atoms can be described by the effective Hamiltonian $\hat H=\tilde\kappa_0\hat p_{ph}\hat p_{a}$, with  $\tilde\kappa_0=a\sqrt{J_xS_x}$, where $a$ is the effective coupling strength \cite{PhysRevLett.85.5639,EPL1}. Obviously, it is a QND-type. An important, immediate application of this Hamiltonian is quantum memory \cite{NATRUE3}. However, initially, such interactions were extensively investigated to produce spin squeezing \cite{EPL1,PhysRevA.60.4974,*PhysRevA.60.2346}.

 We here briefly review the process of spin squeezing based on QND detection. Following Eq. (\ref{equ:0}) the input-output relations for light and atoms can be directly derived as: $\hat x_{ph}^{out}  = \hat x_{ph}^{in}+ \kappa_0\hat p_a^{in} ,\hat p_{ph}^{out}  = \hat p_{ph}^{in},\hat x_a^{out}  = \hat x_a^{in}+ \kappa_0\hat p_{ph}^{in}  ,\hat p_a^{out}  = \hat p_a^{in} $, with $\kappa_0=\tilde{\kappa}_0\sqrt{T}$, where $T$ is the duration of the pulse. Next, a measurement of $\hat x_{ph}^{out}$ is performed, giving a random measurement outcome $\xi$. The momentum operator $\hat p_a^{out}$ is then displaced by an amount $g\xi$, where  $g$ is a gain factor, resulting in $
\hat p_a^{out} = \hat p_a^{in}  - gx_{ph}^{out}  = ( {1 - g\kappa_0 } )\hat p_a^{in}  + g\hat x_{ph}^{in}$. If the light pulse is initially in coherent state such that $\Delta {x_{ph}^{in}}^2=\Delta {p_{ph}^{in}}^2=1/2$, the variance of $\hat p_a^{out}$ can be easily calculated, giving
$(\Delta {\hat p_a^{out} })^2 = \frac{1}
{2}[ {( {1 - g\kappa_0 })^2  + g^2 }].$
Optimizing it, we get $(\Delta {\hat p_a^{out} })^2 =\frac{1}{2}\frac{1}{1+\kappa_0^2}$ for $g=\kappa_0/{(1+\kappa_0^2)}$. Obviously, for nonzero $\kappa_0$, the atomic momentum operator is then squeezed. Finally, we obtain the squeezed spin state (SSS) as:
\begin{eqnarray}
 \hat x_a^{out}=\sqrt{1+\kappa_0^2}\hat x_a^{in},~~~\hat p_a^{out}=\frac{1}{\sqrt{1+\kappa_0^2}}\hat p_a^{in}.\label{equ:0-1}
\end{eqnarray}

\subsection{Controlled-Z gate}
Let us now consider the implementation of optical $C_Z$ gate in an atomic ensemble. Suppose that we have an atomic ensemble prepared in the SSS described above, through which two $x$-polarized light beams are transmitted simultaneously from two perpendicular directions, see Fig. 1(a). For the beam $L$ propagating along the $z$ direction, its interaction with atoms can then be described by $\hat H_L=\tilde{\kappa}_L\hat p_L\hat p_a$. The second beam $M$ propagates along the $y$ direction, leading to the second Hamiltonian $\hat H_M=\tilde\kappa_M\hat p_M\hat x_a$ \cite{PhysRevA.73.022331,PhysRevA.73.062329}. Thus, the complete Hamiltonian for this process can be written as
\begin{equation}
\hat H_1=\tilde\kappa\hat p_L\hat p_a+\tilde\kappa\hat p_M\hat x_a,\label{equ:1}
\end{equation}
where we have assumed $\tilde\kappa_L=\tilde\kappa_M=\tilde\kappa$.
Corresponding to this
Hamiltonian, one may straightly derive the Heisenberg equations for atoms and the Maxwell-Bloch equations (neglecting the effects of light retardation) for light as \cite{PhysRevA.73.022331,PhysRevA.73.062329}:
\begin{equation}
\partial_t\hat x_a \left( t \right) =\tilde\kappa\hat p_L^{in} \left( t \right),\label{equ:2a}
\end{equation}
\begin{equation}
\partial_t\hat p_a \left( t \right) =  - \tilde\kappa\hat p_M^{in} \left( t \right),\label{equ:2b}
\end{equation}
\begin{equation}
 \hat x_L^{out1} \left( t \right) = \hat x_L^{in} \left( t \right) + \tilde\kappa \hat p_a \left( t \right),\label{equ:3a}
 \end{equation}
 \begin{equation}
  \hat p_L^{out1} \left( t \right) = \hat p_L^{in} \left( t \right),\label{equ:3b}
  \end{equation}
  \begin{equation}
  \hat x_M^{out1} \left( t \right) =\hat x_M^{in} \left( t \right) + \tilde\kappa \hat x_a \left( t \right),\label{equ:3c}
  \end{equation}
  \begin{equation}
  \hat p_M^{out1} \left( t \right) = \hat p_M^{in} \left( t \right),\label{equ:3d}
\end{equation}
where $\partial_t$ stands for the partial derivative with respect to $t$. Equations (\ref{equ:2a}) and (\ref{equ:2b}) can be readily solved by integrating over $t$ on both sides, giving
\begin{eqnarray}
  \hat x_a \left( t \right) = \hat x_a \left( 0 \right) + \tilde\kappa\int_0^t {\hat p_L^{in} \left( \tau  \right)} d\tau , \nonumber\\
  \hat p_a \left( t \right) = \hat p_a \left( 0 \right) - \tilde\kappa \int_0^t {\hat p_M^{in} \left( \tau  \right)d\tau } .  \label{equ:4}
\end{eqnarray}
Inserting this set of equations into Eqs. (\ref{equ:3a}) and (\ref{equ:3c}), one will obtain the expressions for $\hat x_{L,M}^{out1}$,
\begin{eqnarray}
  \hat x_L^{out1} \left( t \right) &=& \hat x_L^{in} \left( t \right) +\tilde\kappa \hat p_a \left( 0 \right) \nonumber- \tilde\kappa^2\int_0^t {\hat p_M^{in} \left( \tau  \right)d\tau }, \nonumber\\
  \hat x_M^{out1} \left( t \right) &=&\hat x_M^{in} \left( t \right) + \tilde\kappa \hat x_a \left( 0 \right)+\tilde\kappa^2\int_0^t {\hat p_L^{in} \left( \tau  \right)} d\tau.\label{equ:5}
\end{eqnarray}
Next, we define the dimensionless collective light modes $\hat x_{j}=\int_0^T{f(t)\hat x_{j}(t)dt}$ ($j\in\{L,M\}$) with $[\hat x_{j},\hat p_{j}]=i$, where $f(t)$ is a temporal mode function specifying the mode in question. For the symmetric mode with $f(t)=1/\sqrt{T}$, Eqs. (\ref{equ:5}) are then changed into
\begin{eqnarray}
  \hat {x}_L^{out1}  &=& \hat {x}_L^{in}  + \kappa\hat p_a\left(0\right)- \frac{\kappa^2}{T^{3/2}}\int_0^T {\left(T-t\right)\hat p_M^{in} \left( t \right)dt},  \nonumber\\
  \hat {x}_M^{out1}  &=& \hat {x}_M^{in}  + \kappa\hat x_a\left(0\right)+ \frac{\kappa^2}{T^{3/2}}\int_0^T {\left(T-t\right)\hat p_L^{in}\left(t\right)dt},\nonumber\\
  \label{equ:6}
\end{eqnarray}
where we have interchanged the order of the double integral and defined the dimensionless coupling strength $\kappa=\tilde\kappa\sqrt{T}$. Eqs. (\ref{equ:6}) indicate that, besides the momentum operators $\hat p_M^{in}$ and $\hat p_L^{in}$, the position operators $\hat x_L^{out1}$ and $\hat x_M^{out1}$ also pick up the information of the input atomic operators $\hat p_a(0)$ and $\hat x_a(0)$. Such information bring unfavorable influence on our
scheme, since, for an ideal optical $C_Z$ gate, only the information of $\hat p_M^{in}$ and $\hat p_L^{in}$ are allowed to be admixed into $\hat x_L^{out1}$ and $\hat x_M^{out1}$, respectively.

In the next step, to eliminate the influence of the atomic operators contained in Eqs. (\ref{equ:6}), we propose reflecting the two beams back into the cell after they have completely passed through the atomic ensemble, see Fig. 1(a). Before the second interaction, two delay lines are used to make sure that the two interactions do not overlap in time line, as shown in Fig. 2(b). For the second transmission, beam $M$ still propagates along $y$, leading to the same Hamiltonian as $\hat H_M$, while beam $L$ runs along $-z$ and sees therefore $-\hat x_A$. As a result, its interaction with atoms is changed into $-\hat H_L$. Consequently, the second complete Hamiltonian reads as
\begin{equation}
\hat H_2=-\tilde\kappa\hat p_L\hat p_a+\tilde\kappa\hat p_M\hat x_a.\label{equ:6-2}
  \end{equation}
 From this Hamiltonian, one can easily derive the evolution equations for both light and atoms along the same line outlined above. Using the output
state of the first interaction as the input for the second interaction, we can get the final input-output relations for light,
\begin{eqnarray}
  \hat {x}_L^{out2}  &=& \hat {x}_L^{out1}  - \kappa\hat p_a\left(T\right)+\frac{\kappa^2}{T^{3/2}}\int_0^T {\left(T-t\right)\hat p_M \left( t \right)dt},  \nonumber\\
  \hat {x}_M^{out2}  &=& \hat {x}_M^{out1}  + \kappa\hat x_a\left(T\right)- \frac{\kappa^2}{T^{3/2}}\int_0^T {\left(T-t\right)\hat p_L\left(t\right)dt},\nonumber\\\label{equ:7}
\end{eqnarray}
Inserting Eqs. (\ref{equ:4}) and (\ref{equ:6}) into Eq. (\ref{equ:7}), we finally obtain
\begin{equation}
  \hat {x}_L^{out2}  = \hat {x}_L^{in}  +\kappa^2\hat p_M^{in},\label{equ:8a}
\end{equation}
\begin{equation}
  \hat {p}_L^{out2}  = \hat {p}_L^{in},\label{equ:8b}
\end{equation}
\begin{equation}
  \hat {x}_M^{out2}  = \hat {x}_M^{in}+\kappa^2\hat p_L^{in}  +\frac{2\kappa}{\sqrt{1+\kappa^2_0}}\hat x_a^{in},\label{equ:8c}
\end{equation}
\begin{equation}
  \hat {p}_M^{out2}  = \hat {p}_M^{in},\label{equ:8d}
\end{equation}
where we have substituted the squeezed input atomic state (\ref{equ:0-1}) into Eq. (\ref{equ:8c}) and have assumed that the position quadrature (but not the momentum quadrature) is initially squeezed.
From above one can clearly see that, if the input atomic state is infinitely squeezed (such that $\kappa_0\rightarrow \infty$), then we can neglect the last two terms of  Eq. (\ref{equ:8c}). In this case, if $\kappa$ is equal to $1$, then an ideal $C_Z$ gate between light $L$ and $M$ is performed.

However, in reality, the strength of coupling of light to atoms is always finite. As a result, the noise terms contained in Eq. (\ref{equ:8c}) can not be completely eliminated. In this case, we need to quantify the performance of the $C_Z$ operation. Often, it can be done by calculating the fidelity $F=\langle\psi|\hat\rho_{out}|\psi\rangle$, which is a measure of how well the output
state $\hat\rho_{out}$ compare to the original input state $|\psi\rangle$. Besides, we note that both the atomic state and the light states involved here are all Gaussian. For an $N$-modes Gaussian state, the Wigner function can be conveniently expressed in the form \cite{RevModPhys.77.513,arxiv1}: $W=1/(\pi^N\sqrt{\text{det}\gamma})\text{exp}[-(\xi-\textbf{m})\gamma^{-1}(\xi-\textbf{m})^T]$, where $\gamma$ stands for the covariance matrix, and $\xi$ denotes the $N$-dimensional vector having the quadrature
pairs of all $N$ modes as its components, while $\textbf{m}$ represents the mean values. Mathematically, the fidelity $F$
can be calculated by the overlap of the pure input state $W_{in}$ with the mixed output state $W_{out}$
\begin{eqnarray}
 F &=&{\left(2\pi\right)}^N\int_{ - \infty }^{ + \infty } {dx} \int_{ - \infty }^{ + \infty } {dp} {W_{in}}\left( {x,p} \right){W_{out}}\left( {x,p} \right) \nonumber\\
  &=& \frac{2^N}{{{\pi ^N}\sqrt {\det \gamma \det \gamma '} }}\int_{ - \infty }^{ + \infty } {dx} \int_{ - \infty }^{ + \infty } {dp} {e^{ - \left( {\xi  - \textbf{m}} \right){\gamma ^{ - 1}}{{\left( {\xi  - \textbf{m}} \right)}^T}}} \nonumber\\
  &&\times {e^{ - \left( {\xi  - \textbf{m}'} \right){{\gamma '}^{ - 1}}{{\left( {\xi  - \textbf{m}'} \right)}^T}}} \nonumber\\
  &=& 2^N\sqrt {\det{{{\left( {\gamma  + \gamma '} \right)}^{ - 1}}}} {e^{ - \left( {\textbf{m} - \textbf{m}'} \right){{\left( {\gamma  + \gamma '} \right)}^{ - 1}}{{\left( {\textbf{m} - \textbf{m}'} \right)}^T}}},\label{equ:9}
\end{eqnarray}
where the output state is characterized by the mean value $\textbf{m}'$ and the covariance matrix $\gamma'$. In the present case, we have $N=2$, and, if we assume the input variables $\hat x_a^{in}$ and $\hat p_{ph}^{in}$ are centered around zero such that $\textbf{m}=\textbf{m}'=0$, then the average values of the output states in Eqs. (\ref{equ:8a})$-$(\ref{equ:8d}) are conserved. Consequently, the fidelity of Eq. (\ref{equ:9}) can be further simplified as $F=4\sqrt{\text{det}(\gamma+\gamma')^{-1}}$. For the case of cluster state creation, the two beams before entering the $C_Z$ are always prepared in a squeezed vacuum state \cite{PhysRevA.73.032318,PhysRevLett.104.250503}. Hence, without loss of generality, we can assume that the two input light modes have a normalized variance $(\Delta\hat x_{L,M}^{in})^2=1/(\Delta\hat p_{L,M}^{in})^2=\frac{1}{2}e^{-2s}$ with $s$ the squeezing parameter. With this assumption, we finally obtain the fidelity as:
\begin{equation}
F = \frac{1}{{\sqrt {1 + \frac{2e^{2s}}{{1 + \kappa _0^2}}} }}\label{equ:10},
\end{equation}
 \begin{figure}[tp]
\centering
\includegraphics[scale=0.65]{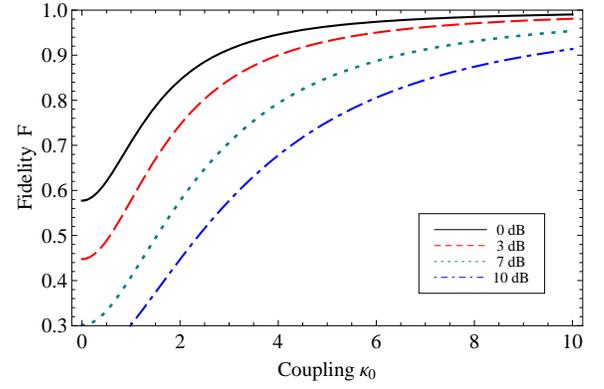}
\caption{(Color online) The fidelity of the gate operation vs the coupling strength $\kappa_0$ for different squeezing of the input light. The solid line denotes the coherent input.}
\end{figure}
where we have put $\kappa=1$. Corresponding to this expression, in Fig. 2, we are
able to show the fidelity of the gate operation in its dependence
on the coupling strength $\kappa_0$ for different squeezing of the input light (in dB). As can be seen from the figure that large coupling strength are required for the achievement of high fidelities.
 \section{noise effects and experimental feasibility}
 \subsection{Noise effects}
So far, we have neglected the noise effects. As in reality,
atoms are usually contained in glass cells. Therefore, light reflections by the cell walls become inevitable. Such reflections, however, can be modeled by a beam splitter type admixture of vacuum components \cite{PhysRevA.73.022331,PhysRevA.72.052313}, which transforms
the input light quadrature as $\hat \vartheta^{in}\to\hat \vartheta^{in'}=\sqrt{1-r}\hat \vartheta^{in}+\sqrt{r}\hat \vartheta^v$ with $\hat\vartheta\in\{\hat x_{L,M},\hat p_{L,M}\}$, where $r$ is the reflection coefficient and $\hat\vartheta^v$ is the vacuum noise quadrature. On the other hand, due to the weak excitation by the light
beams, the atoms also undergo dissipation. We assume that
the spontaneous emission happens at a rate of $\eta/T$ \cite{PhysRevA.73.062329}.
With these modeling, the evolution equations (\ref{equ:2a})$-$(\ref{equ:3d}) are then changed into
\begin{equation}
 {\partial _t}{{\hat x}_a}\left( t \right) =\tilde \kappa' \hat p_L^{in}\left( t \right) - \frac{\eta }{{2T}}{{\hat x}_a}\left( t \right) + \sqrt {\frac{\eta }{T}} \hat x_a^v\left( t \right), \nonumber\\
 \end{equation}
 \begin{equation}
 {\partial _t}{{\hat p}_a}\left( t \right) =  - \tilde \kappa ' \hat p_M^{in}\left( t \right) - \frac{\eta }{{2T}}{{\hat p}_a}\left( t \right) + \sqrt {\frac{\eta }{T}} \hat p_a^v\left( t \right), \nonumber \\
  \end{equation}
 \begin{equation}
 \hat x_L^{out1}\left( t \right) =\hat x_L^{in'}\left( t \right) + \tilde \kappa' {{\hat p}_a}\left( t \right), \nonumber \\
  \end{equation}
 \begin{equation}
 \hat p_L^{out1}\left( t \right) =\hat p_L^{in'}\left( t \right),  \nonumber\\
   \end{equation}
  \begin{equation}
 \hat x_M^{out1}\left( t \right) = \hat x_M^{in'}\left( t \right) +\tilde \kappa' {{\hat x}_a}\left( t \right),  \nonumber\\
  \end{equation}
 \begin{equation}
 \hat p_M^{out1}\left( t \right) = \hat p_M^{in'}\left( t \right),
 \end{equation}
where $\hat x_a^v$ and $\hat p_a^v$ are Langevin noise operators with zero
mean, having $\langle \hat x_a^v(t)\hat p_a^v(t')\rangle=\delta(t-t')/2$. Here, we have defined the reduced coupling strength $\tilde\kappa'=\sqrt{1-r}\tilde\kappa$. Corresponding to this set of equations, one may derive the modified input-output relations for light
\begin{eqnarray}
 \hat x_L^{out1} &=& \hat x_L^{in'} + \kappa '{{\hat p}_a}\left( 0 \right)
  - \frac{{{{\kappa '}^2}}}{{{T^{3/2}}}}\int_0^T {dt} \nonumber\\
 &&\times{e^{ - \frac{{\eta t}}{{2T}}}}\left( {T - t} \right)\left[ {\hat p_M^{in'}\left( t \right) - \frac{\sqrt {\eta}}{{\kappa '}}\hat p_a^v\left( t \right)} \right],\nonumber\\
 \hat x_M^{out1} &=& \hat x_M^{in'} + \kappa '{{\hat x}_a}\left( 0 \right)
  + \frac{{{{\kappa '}^2}}}{{{T^{3/2}}}}\int_0^T {dt}\nonumber\\
  &&\times{e^{ - \frac{{\eta t}}{{2T}}}}\left( {T - t} \right)\left[ {\hat p_L^{in'}\left( t \right) + \frac{\sqrt {\eta}}{{\kappa '}}\hat x_a^v\left( t \right)} \right]\label{equ:11},
 \end{eqnarray}
and for atoms
\begin{eqnarray}
 {{\hat x}_a}\left( T \right) &=& {e^{ - \frac{\eta }{2}}}{{\hat x}_a}\left( 0 \right) + \frac{1}{\sqrt T}\int_0^T {dt} \nonumber\\
  &&\times{e^{ - \frac{{\eta t}}{{2T}}}}\left[ {\kappa '\hat p_L^{in'}\left( t \right) + \sqrt {\eta} \hat x_a^v\left( t \right)} \right],\nonumber\\
 {{\hat p}_a}\left( T \right) &=& {e^{ - \frac{\eta }{2}}}{{\hat p}_a}\left( 0 \right) - \frac{1}{\sqrt T}\int_0^T {dt}\label{equ:12}\nonumber\\
 &&{e^{ - \frac{{\eta t}}{{2T}}}}\left[ {\kappa '\hat p_M^{in'}\left( t \right) - \sqrt {\eta} \hat p_a^v\left( t \right)} \right].
 \end{eqnarray}
Before reflecting back into the vapor, the two light pulses will experience another two crossing of the cell wall [see Fig. 1(a)], which transfers the light states of Eqs. (\ref{equ:11}) into $\hat x_i^{out1}\to\hat x_i^{out1'}=\sqrt{1-2r}\hat x_i^{out1}+\sqrt{2r}\hat x_i^v$ with $i\in\{L,M\}$. Using the light state $\hat x_i^{out1'}$ and the atomic state of Eqs. (\ref{equ:12}) as the input states, the second interaction occurs, resulting in the final output states
\begin{eqnarray}
 \hat x_L^{out2} &=& \sqrt {1 - 3r} \left[ {\hat x_L^{in} + \epsilon _1^ - {\kappa ^2}\hat p_M^{in}} \right.\nonumber\\
&&\left. { + \epsilon _2^ - {\kappa ^2}\hat{\tilde{p}}_M^{in} + \frac{\eta }{2}\kappa {{\hat p}_a}\left( 0 \right)} \right] + {{\hat f}_{v1}},\label{equ:13a}\nonumber\\
\hat p_L^{out2} &=& \sqrt {1 - 3r} \hat p_L^{in} + \sqrt {3r} \hat p_L^v,\label{equ:13b} \nonumber\\
x_M^{out2} &=& \sqrt {1 - 3r} \left[ {\hat x_M^{in} + \epsilon _1^ + {\kappa ^2}\hat p_L^{in}} \right. \nonumber\\
 &&\left. { + \epsilon _2^ + {\kappa ^2}\hat {\tilde p}_L^{in} + 2\kappa \left( {1 - \frac{\eta }{4}} \right){{\hat x}_a}\left( 0 \right)} \right] + {{\hat f}_{v2}}, \label{equ:13c} \nonumber\\
 \hat p_M^{out2} &=& \sqrt {1 - 3r} \hat p_M^{in} + \sqrt {3r} \hat p_M^v,\label{equ:13d}
 \end{eqnarray}
  \begin{figure}[tp]
\centering
\includegraphics[scale=0.7]{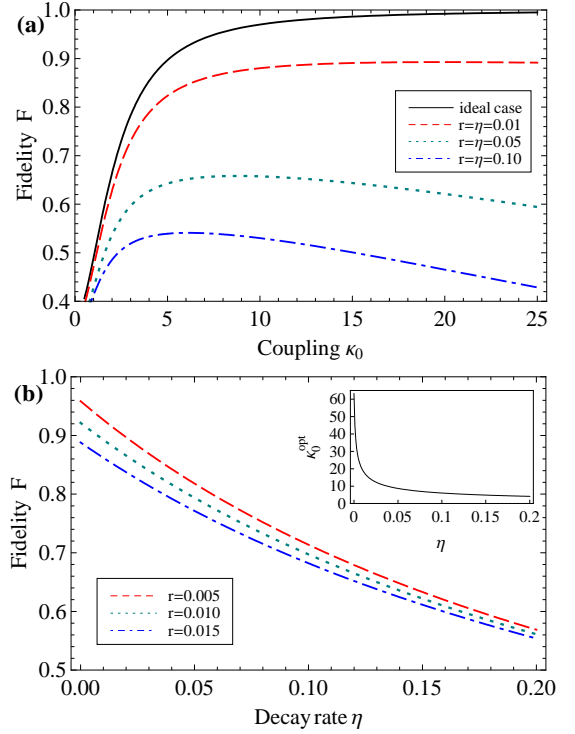}
\caption{(Color online) (a) Gate operation fidelity $F$ vs coupling $\kappa_0$ in the
presence of atomic decay and light reflection. The input light states
are assumed to be polarization squeezed with the degree of $5$ dB. (b) Optimized fidelity vs decay rate
$\eta$ for different  reflection coefficient. The inset shows how the
optimal coupling $\kappa_0$ varies with $\eta$.}
\end{figure}
where $\epsilon_1^{\pm}=(1-r)(1 \pm 2r)( {1 - \frac{\eta }{2} \mp \frac{{2r}}{{1 \pm 2r}}})$ and
$\epsilon_2^{\pm}=(1-r)\frac{{(1 \pm 2r)}}{{\sqrt 3 }}( {\frac{\eta }{2} \pm \frac{{2r}}{{1 \pm 2r}}} )$, and we defined the total vacuum noise $\hat f_{v1}$ and $\hat f_{v2}$, which are a function of the vacuum operators $\hat x_i^v$ and $\hat p_{i}^v$, with $i\in\{a,L,M\}$. Here, $\hat{\tilde p}_{L,M}$ are new defined collective light modes with a temporal mode function $f(t)=(1-2t/T)$, satisfying $[ \hat{\tilde x}_j,\hat{\tilde p}_j ]=i$ with $j\in\{L,M\}$. It is easily checked that they are independent from all other modes \cite{PhysRevA.72.052313}.
Unlike the ideal case, Eqs. (\ref{equ:13a}) show that, besides the atomic position quadrature, the momentum quadrature $\hat p_a(0)$ now also appears. This term arising because of the decoherence of atoms, as we shall see, will lead to the optimal implementation of the scheme. Finally, after taking the end reflection losses into account (because of the fourth crossing of a cell wall) by
damping Eqs. (\ref{equ:13a}) with a factor
$\sqrt{1-r}$ and adding appropriate noise terms, the covariance matrix of the final output state (and, thus, the fidelity) can
then be calculated directly. Putting $\kappa=1$, the fidelity versus coupling strength $\kappa_0$ for squeezed input light with $s=\frac{1}{4}\text{ln}~10$ (corresponding to $5$ dB) in the presence of noises
is depicted in Fig. 3(a) for parameters $r=\eta=0.01$, $r=\eta=0.05$, and $r=\eta=0.1$. In each case, there exist an optimal fidelity, $F=0.89$, $F=0.66$, and $F=0.54$, for $\kappa_0=19.90$, $\kappa_0=8.78$, and $\kappa_0=6.12$, respectively. As illustrated by the
graph, decay and reflections losses have a significant effect on the quality of the gate operation. Losses of the
latter kind, however, can be greatly reduced down (to about $0.5\%$) with improved antireflection coating \cite{PhysRevA.74.064301}. Figure 3(b) shows the $\kappa_0$-optimized fidelity
versus $\eta$ for different (small) values of the reflection parameter $r$. For $\eta=0.1$ (corresponding to the experimental conditions of \cite{NATRUE8}) and $r=0.005$, a fidelity $F=0.71$ would be possible corresponding to $\kappa_0=6.05$.

\subsection{Experimental feasibility}
To successfully and efficiently implement the
$C_Z$ gate, it is required that (i) $T<T_{DL}$, where $T_{DL}$ represents a time in which the two beams
pass through the delay lines, (ii) the coupling strength $\kappa=1$, and (iii) large interaction strength $\kappa_0$ is achievable. Conditions (i), as analyzed in Ref. \cite{PhysRevA.78.010307}, is feasible within presently established techniques, i.e. by using a 1 $\mu s$ delay line \cite{phd2} and a sub-$\mu s$ pulse \cite{ph3}. Condition (ii) has been realized in many physical systems, e.g. room temperature atomic vapors \cite{NATRUE3}. Condition (iii), however, is by now still experimental challenge. Although, theoretically, one can always enhance the coupling strength by increasing the intensity of the light beams, or, the density of atoms, such enhancement, on the other hand, will cause high decay rate, and thus will lower the efficiency of current scheme. To overcome this limitation, we propose to inject squeezed light instead of coherent light during the spin squeezing process, which transfers the input quadratures $(\hat x_{ph}^{in},\hat p_{ph}^{in})$ into $(e^{-s}x_{ph}^{in},e^{s}\hat p_{ph}^{in})$. With this setting, the spin state of Eqs. (\ref{equ:0-1}) is then changed into
\begin{eqnarray}
\hat x_a^{out}=\sqrt{1+e^{2s}\kappa_0^2}\hat x_a^{in},~\hat p_a^{out}=\frac{1}{\sqrt{1+e^{2s}\kappa_0^2}}\hat p_a^{in}.
\end{eqnarray}
Note that the coupling strength is now enhanced by a factor $e^s$. We can define the effective coupling strength $\kappa^{eff}_0=e^s\kappa_0$. As a result, the higher the degree of squeezing is, the larger the effective coupling strength becomes. For room temperature atomic vapors, the feasible value of coupling strength is around $\kappa_0\approx1.4$ \cite{phd1}. In this case, if we inject a light with $8.5$ dB of squeezing, a large coupling strength $\kappa^{eff}_0\approx10$ can then be achieved.

\section{conclusions}
In conclusion, we have presented a simple and realistic scheme for realizing CV $C_Z$ gate in an atomic ensemble. The process is based on off-resonant interaction between light and spin-polarized
atomic ensembles. By sending two off-resonant pulses propagating in two orthogonal directions twice through an atomic ensemble which is initially prepared in spin squeezed state, we find that a $C_Z$ operation between the two pulses is performed. The more the amount of spin squeezing input, the higher the fidelity we will obtain. We also considered the influences of the noise effects including the atomic decay and photon reflections by the cell walls, showing that they have a strong effect on the fidelity. Noises of later kind, however, can be greatly suppressed by adding antireflection coating to cell walls. Such suppressions enable us to achieve quite high fidelities with current experimental parameters.
It is well known that offline squeezing and $C_Z$ gate together enable the construction of arbitrarily large CV cluster states \cite{PhysRevA.79.062318,PhysRevLett.104.250503}. Recently, offline squeezing based on atomic ensembles has been proposed by Sherson \emph{et al.} \cite{PhysRevLett.97.143602}. Therefore, our proposal paves the way for the implementation of CV one-way QC based only on atomic ensembles.

\begin{acknowledgments}
This work was supported by the Natural Science Foundation
of China (Grants No. 11074190 and No. 10947017), and the Natural Science Foundation of Zhejiang province, China (Grant No. Y6090529).
\end{acknowledgments}
\bibliographystyle{apsrev4-1}
\bibliography{myreference}

\end{document}